# Why it is Worth Making An Effort when Learning With GenAI[1]

Yvonne Rogers
UCLIC, UCL, UK

**Abstract**

Students routinely use ChatGPT and the like now to help them with their homework, such as writing an essay. It takes less effort to complete and is easier to do than 'by hand'. It can even produce as good if not better output than the student's own work. However, there is a growing concern that over-reliance on using GenAI in this way will stifle the development of learning writing and critical thinking skills. How might this trend be reversed? What if students were required to make more effort when using GenAI to do their homework? It might be more challenging, but the additional effort involved could result in them learning more and having a greater sense of achievement. This tension can be viewed as a form of 'effort paradox'; where effort is both viewed as something to be avoided but at the same time is valued. Is it possible to let students learn sometimes with less and other times more effort? Students are already adept at the former but what about the latter? Could we design new kinds of AI tools that deliberately require more effort to use to deepen the learning experience? In this paper, I begin to outline what form these might take, for example, asking students to use a combination of GenAI tools with traditional learning approaches (e.g. note-taking while reading). I also discuss how else to design tools to think with that augments human cognition; where students learn more the skills of metacognition and reflection.

**Introduction**

Millions of children are routinely using ChatGPT and other generative AI tools in their everyday lives and at school. Examples include planning an essay, writing a poem, creating a podcast, generating images and finding activities to do when travelling. They are also using GenAI in imaginative ways, for example, to rewrite the lyrics of Taylor Swift's "Shake It Off" through the eyes of a squirrel (Montgomery, 2025). In 2024 the National Literacy Trust reported over 50% of young people aged between 13 to 18 years said that GenAI had helped them come up with new ideas, understand things, help with their reading and to learn new things. Only a year later, another report found that as many as 92% of students were now using GenAI (Freeman, 2025). For many children, ChatGPT has replaced Google as their go to tool when wanting to learn or find something out.

What are the implications of this widespread uptake of GenAI for education? On the one hand, it has much potential for transforming learning, for example, to be more creative, engaging and self-initiated. Students can be asked to spend time reflecting on how good ChatGPT's answer is to the questions they pose and how they can improve upon it. Instead of simply regurgitating what they find on Wikipedia or other online resources they can be asked to develop and hone their critical and analytical skills. In so doing, they can learn what makes for a good or poor argument, developing metacognition skills in the process. Teachers can also use AI, themselves, to rethink more challenging assignments to set while providing extensive forms of interactive feedback that students can follow up with. For example, teachers could ask GenAI to suggest new topics to explore and complex problems to solve (Kaneski *et al.,* 2023) which they could try doing themselves first using various GenAI tools.

On the other hand, there has been much debate about the potential downsides of students becoming overly reliant on GenAI especially for tasks that require critical thinking and problem-solving. There is concern that their widespread use could reduce opportunities to learn and practice various literacy skills,

---

such as writing, researching and summarising. In the long-term, this could impact students' cognitive development, especially their ability to think and express their thoughts, since learning to think is inextricably bound with learning to write (Ackerman, 1993). It has been argued that if students no longer need to write they will spend less time thinking things through, that could result in their thoughts being less clear or cohesive (Reich, 2023).

So, what can be done to mitigate the potential risks while enabling students to benefit from using AI tools? Can students be encouraged to think critically when using AI tools during learning tasks rather than simply accepting what they suggest? Is it possible to develop AI-enabled tools or adapt existing ones that require students sometimes to make 'more effort' while other times accept that it is not a problem that they use them to make 'less effort'? To understand better how we might achieve a balance we consider the effort paradox.

**The effort paradox**
Many well-known models in cognitive psychology, neuroscience, and economics have suggested that effort is costly. In a nutshell, people tend to avoid making an effort, where possible, when given a choice. For example, many people prefer to have their weekly shop or takeaway food delivered to their homes rather than travel to and from a store to buy it. It requires much less effort, even though it is often more expensive. Inzlicht *et al.* (2018), however, suggest that the opposite may be true – where making an effort can actually be rewarding. So, people who make the effort to go to a store and buy ingredients and then make a meal at home find it more rewarding than having a meal delivered. This is known as the *effort paradox*. Another way of viewing this is as a form of the so-called *IKEA effect*. This term was coined to describe how making a piece of furniture from a flatpack kit can be more rewarding than when it arrives preassembled - even though it takes much more time and effort (Norton *et al.*, 2012).

The effort paradox and the IKEA effect can be useful concepts for exploring how Gen AI is transforming learning, for better or for worse. As is well known, many people who frequently use GenAI to help them with their everyday tasks have discovered how it reduces the time and effort to do what they did beforehand. The same holds true for students when learning and doing their homework. In many contexts saving time in this way can be beneficial. For example, many students now use GenAI to help them overcome the so-called 'blank page' problem, especially those whose English is not their second language, and who struggle to know how to begin an essay or a report. Asking GenAI to help them get started can give them idea and make a plan from which they can add to, agree or disagree with. Similarly, it can help students who have muddled thoughts help provide a way of structuring them in a more ordered way.

But at the same time, could GenAI tools be designed that reward students if they make more effort? In particular, could they encourage them to make *more* effort when learning, such that they see the benefit of doing so, and from this gain a sense of achievement?

**Designing new GenAI tools to enhance learning and thinking**
There are a number of ways we can think of how to design and use AI tools to facilitate learning that take more effort. Here I suggest five that we have been working on:
   (i)   Develop GenAI 'tools for thought' that deliberately provoke critical thinking, provide personalized tutoring, and enable novel ways of sensemaking (Tankelevitch *et al.,* 2025).
   (ii)  Use GenAI tools to slow down student thinking in order to get them to reflect more on what they are thinking about. For example, they could be asked to complete a learning task by following a series of steps, sometime using AI and other times using other kinds of tools or just by themselves. For this, an AI assistant could be created to ask a series of open-ended questions while setting personalised challenges they have to work through.
   (iii) Externalise student's half-baked ideas and muddled thoughts through interacting with GenAI in an iterative way until they make more sense and become clearer.
   (iv)  Constrain the GenAI tools and tasks to enable students to conceptualise problems they are trying to solve from different perspectives (Yuill and Rogers, 2012). An example of a

|     | constraint is the use of software scaffolding where the student is guided to follow a series of steps (e.g. voice prompts). |
| --- | --- |
| (v) | Design the GenAI to help students think for themselves (Rogers, 2022, 2024) by developing new 'supertools' that can expand their minds, daring them to think differently, while at the same time helping them break through the barriers that often stall or prevent creative leaps. For example, the GenAI could be designed to augment what they do by suggesting, prompting, conjuring up, counter-arguing, nudging, probing and even acting as a sparring partner or a form of super-ego. |

To this end, my colleagues and I have started developing virtual agents and AI tools that can scaffold teams and individuals during various sensemaking, decision-making and learning tasks. To begin, we developed chatbots that are proactive, called Proberbots (Reicherts *et al.,* 2022b) to intervene at opportune moments during learning and decision-making tasks. They do so, by questioning and nudging people to think in a different direction. In an early study, we designed a Proberbot called VoiceViz that appeared at the interface when teams appeared stuck during a reasoning task prompting them to change tact or consider a different way of solving the problem (Reicherts et *al.,* 2022a). The task they were asked to do was to analyse and explain a set of complex health data that was represented as a series of visualisations. VoiceViz was programmed to ask open-ended questions, such as: *"Do you need a hint for analysing X?", "Did you consider the difference between variable X and Y?"* or *"What might have caused the sudden spike?"* The findings from our study showed how this Proberbot engendered more collaborative interactions, leading to team members having deeper discussions, involving new trains of thought while enabling them to generate more hypotheses.

Following this initial work on designing chatbots to elicit reflection and new ways of thinking, we developed another kind of software scaffold, called SelVReflect. This time, we combined a voice-based assistant with a VR tool with the aim of encouraging people to immerse themselves in and reflect on their past challenges (Wagener *et al.,* 2023). 20 participants were asked to use the tool to approach a personal challenge and its (emotional) components from different perspectives and to discover new relationships between these. The software constraints that were introduced were different kinds of voice prompts. The findings from the study showed the participants developed a better understanding of the situation and of themselves when using the voice/VR tool. Hence, both VoiceViz and SelVReflect were able to probe people's thinking, facilitate cognitive externalisation and foster reflection.

More recently, we have begun exploring how to embed GenAI in existing software tools that encourages users to think more about their goals and intentions when making decisions. We designed AI tools that helped them work through hunches that were poorly formulated while also helping reduce impulsivity (Reicherts *et al.,* 2025). In one study, we developed two kinds of AI to help people make investment decisions, which are often highly complex and open-ended. For this, we implemented a simulated stock trading platform in which we embedded an AI investing 'assistant' offering two different types of assistance: one that makes direct recommendations to users as to what they might do (called RecommendAI) and the other that requires users to describe their own rationale first before making their choice (called ExtendAI), which the AI assistant elaborated on by embedding its feedback, including suggestions for what else they might do, into it.

We conducted a study comparing how 20 participants used the two different AI assistants and found that ExtendAI integrated better into the decision-making process and users' own thinking leading to better outcomes. There was a cost, however, for using ExtendAI which was that the participants had to write down their own rationale first, which they found burdensome. On the other hand, they mentioned how making this extra effort was valued. Unexpectedly, participants commented on how they liked RecommendAI as it provided more novel insights than they would have thought of by themselves. This led them to consider which of these might be better options, getting them to reflect at a different stage of the decision-making. Hence, our study showed that GenAI tools that require more or less effort can have different benefits for decision-making. Having a suggestion made by RecommendAI, rather than having to come up with one themselves, helped participants think about other possibilities they might not have otherwise. The effort involved required evaluating the suggestions made by the AI at the end of the task to determine if they were useful compared with upfront effort required to generate rationale for a decision when using ExtendAI which led to them being more reflective in their own thinking.

Extrapolating from these findings into the domain of learning suggests that it can be beneficial for students to understand how, when and where to make more or less cognitive effort in their learning tasks, with and without GenAI, and the outcomes of doing so. For example, they can ask the AI to help them get started on a task (e.g. an essay), and in doing so use less effort. Later on, they can be expected to expend more cognitive effort, when required to evaluate the outcome of the GenAI suggestions, critiquing and checking the validity of the this, and iterating their prompts on what has been said or suggested. The goal is for them to to see the benefits of how making an effort to hone their thinking skills in this way can result in a sense of accomplishment. The same with a coding exercise – less effort at the beginning to write prompts to generate code and then more effort later to check, tweak and critique the GenAI output.

Hence, the idea here is to make students more aware of the value of making an effort when learning. The question remains as to whether increasing their awareness in this way will help them learn better and remember more. In particular, can they develop metacognitive skills so that they know how best to use GenAI and other software tools to optimise a sense of achievement? Shifting the amount of effort to different stages of a task when using GenAI tools, has been discussed by Lui *et al.* (2025). They conducted a survey of how knowledge workers are changing their workflow habits through their routine uses of GenAI. They found this happened in three different ways: (i) from information gathering to information verification, (ii) from problem-solving to AI response integration; and (iii) from task execution to task stewardship. In particular, the survey respondents commented on how they took less effort in retrieving and curating task-relevant information but, conversely, more effort for the end of their tasks, when they started checking whether the information the GenAI had generated was correct. Moreover, they found that using GenAI resulted in the knowledge workers creating new cognitive tasks such as assessing the AI-generated content to determine its relevance and applicability to their specific tasks which they would not necessarily do if they had generated the content themselves.

**Other opportunities for learning when using GenAI tools**
Alongside considering how different levels of cognitive effort impact learning when interacting with GenAI we should look at how it can facilitate new learning activities. More generally, Sharples (2023) has suggested reframing AI and education in terms of the social process of conversation and exploration. He calls this social generative AI for education where humans and GenAI agents engage in a broad range of social interactions and where the effort is dynamically shifted between the student and the AI. This conceptualisation leads to asking new research questions, such as how can GenAI be designed to enable humans to engage in conversations when learning and what form should the dialogue take between the AI and the human? He proposes asking children to work together with GenAI to create a story that represents a diversity of views, ideas and perspectives. The rationale is for the students to start by agreeing on a plot and setting for the story, then prompt ChatGPT to generate the opening paragraphs. After reading this they follow by proposing characters and actions, asking ChatGPT to generate different versions that encourage diversity.

We can also rethink how best to assess students' work when using GenAI. Educators could explicitly ask their students to think more about what the AI tools generate based on their prompts. In so doing, they could learn how to think more critically about the 'conversation' they have for the different stages of their assignments and projects, documenting what was their ideas, what they asked, what they were disappointed with, what they disagreed with and what they accepted and how they iterated from this. Similar to Sharples (2023), Tang *et al.* (2024) discuss the pedagogical potential of GenAI, by reconceptualising its role more as a 'dialogic agent', focussing on critical questioning. They conducted a study using GenAI in secondary schools in Australia, where the students were asked to research (i) the historical and social context surrounding a poem of their choice and (ii) ask ChatGPT to take on the persona of a Shakespearean character in order to answer their questions for that character. Their findings showed that students actively engaged in dialogues with GenAI, where follow-up questions suggested critical thinking and creativity had been engendered. The teachers who took part guided his students through a structured process of engaging AI-generated responses in a dialogue by posing follow-up questions to continue the conversation. The students were also encouraged to verify the accuracy of ChatGPT's responses through follow-up research.

GenAI also has much potential for helping students with learning difficulties, for example, by clarifying writing instructions, establishing writing goals, and creating structured writing plans (Onufer, 2024). A recent survey by Zhao *et al.* (2025) of students with neurodiversity, including those with ADHD, specific learning difficulties, and autism, found that that ChatGPT helped them overcome a wide range of academic writing challenges and anxieties, through explaining complex topics, structuring their ideas for writing, improving their word choice and summarising texts and other learning materials.

AI can also be combined with more traditional tools to augment learning. Kreijkes *et al.* (2024) for example, compared using LLMs with traditional note-taking with just notetaking or just using LLMs to complete comprehension and retention tasks. They found 14-15 years olds valued LLMs for making complex material more accessible and reducing cognitive load, but that making notes, themselves without the AI, required more cognitive effort but resulted in deeper engagement while enhancing their memory of the content. Moreover, their findings showed that using LLMs can facilitate initial understanding and student interest in the topic but that making the effort to write their own notes helped them understand and remember more.

**Summary**
When integrating GenAI into learning tasks, it is useful to consider the balance between (i) how much cognitive effort students are prepared to make at different stages of a learning task (especially if they know they can simply ask ChatGPT to do it) and (ii) their willingness to expend more cognitive effort at certain times, such as when evaluating, building upon, iterating or critiquing the output from GenAI. Thinking about how best to spend their cognitive effort for different parts of a learning task may prove to be an effective strategy. For example, spending less effort at the beginning of a task by using GenAI tools can help students get started; at other times, they may be willing to spend more effort themselves, honing their thinking skills, for example, when critiquing and reflecting on what the GenAI has produced. Letting students use GenAI tools (e.g. to generate a plan) alongside traditional learning skills (e.g. note-taking) may also prove to be an effective way for students to develop critical thinking and metacognitive skills.

In sum, there is much potential for appropriating and designing new GenAI tools that can be beneficial for student learning, for example, facilitating debate and discussion that enables students to learn to think more critically. They can also be a great leveller on the playing field, supporting and scaffolding all manner of students in their learning. By using the effort paradox and the IKEA effect I have shown how we can view the future of learning and AI. There are of course many other ways to conceptualise the ever-evolving relationship between learning and AI, especially those that take into account pedagogical theories. Here, I have tried to show how we can rethink how to combine and constrain GenAI tools to encourage students to think more for themselves and in doing so, be able to reflect on their learning – without it being too much effort all the time while enabling them to feel a sense of accomplishment when they do make an effort.

**Acknowledgements**

I would like to thank James Katz, Leon Reicherts, Ava Scott and Abi Sellen for their comments on an earlier draft.